\documentclass[12pt]{article}

\usepackage{amsthm}
\usepackage{amssymb}
\usepackage{authblk}
\usepackage{dsfont}
\usepackage{colortbl,dcolumn}
\usepackage{color}
\usepackage{indentfirst}

\usepackage{booktabs}
\usepackage{setspace}

\usepackage{amsfonts,amsmath}
\usepackage{hyperref}
\usepackage{graphicx}
\usepackage{caption}
\usepackage{subcaption}

\makeatletter
\g@addto@macro\normalsize{%
}
\makeatother

\newcommand{\beq}[1]{\begin{equation} \label{#1}}
\newcommand{\eeq}{\end{equation}}
\newcommand{\bed}{\begin{displaymath}}
\newcommand{\eed}{\end{displaymath}}
\newcommand{\bea}{\bed\begin{array}{rl}}
\newcommand{\eea}{\end{array}\eed}

\newcommand{\barray}{\begin{array}{ll}}
\newcommand{\earray}{\end{array}}
\def\({\left(}
\def\){\right)}

\date{}

\begin{document}
\onehalfspacing 

\title{How Does Monetary Policy Influence the U.S. Treasury Bond Yields, and What are the Implications for Portfolio Managers?
\footnotemark[2]
\footnotetext[2]{E-mail addresses: 
kz2090@nyu.edu,
lyuhanm@umich.edu, 2210503016@stu.suda.edu.cn.
}
}

\author[a]{Minnie Zhu}
\author[b]{Yuhan Liu}
\author[c]{Simon Gong}
\affil[a]{
\small Department of Economics, New York University, New York, NY 10012}
\affil[b]{
\small Department of Mathematics, University of Michigan, Ann Arbor, MI 48109}
\affil[c]{
\small School of Business, Soochow University, Su Zhou, Jiangsu, China, 215006}

\maketitle

\begin{abstract}
This paper investigates the impact of monetary policy surprises on U.S. Treasury bond yields and the implications for portfolio managers. Based on the supply and demand model, traditional economic theories suggest that Federal Reserve bond purchases should increase bond prices and decrease yields. However, New Keynesian models challenge this view, proposing that bond prices should not necessarily rise due to future expectations influencing investor behavior. By analyzing the effects of monetary policy surprises within narrow windows around Federal Open Market Committee (FOMC) announcements, this study aims to isolate the true impact of these surprises on bond yields. The research covers Treasury bonds of various maturities--3 months, 1 year, 10 years, and 30 years--and utilizes cross-sectional regression analysis. The findings reveal that financial crises significantly decrease short-term yields, while no obvious evidence of factors that might affect long-term yields. This paper provides insights into how monetary policy influences bond yields and offers practical implications for portfolio management, particularly during quantitative easing and financial crises. 
\\
	
\end{abstract}

\section{Introduction}
Portfolio managers manage interest rate risk\footnote{The risk that happens when changes in interest rates negatively affect the value of their bond holdings.}. Understanding the relationship between monetary policy and bond yields is crucial for them as it allows managers to hedge against this risk. For instance, if managers expect an increase in interest rates, they might reduce the duration of their portfolios by shifting to short-term bonds to minimize losses (see \cite{RD14}). Bond yields, as well, can reflect inflation expectations and forecast economic growth. Central banks use rate changes to influence economic activity through different monetary policy transmission mechanisms. Provided portfolio managers understand these dynamics, they can better anticipate central bank actions and their potential influence on financial markets, helping them earn higher profits (\cite{Ben13}).

Traditional economic models, such as the supply and demand model, predict that Federal Reserve bond purchases should raise bond prices. When the bond prices rise, the bond yield decreases. However, New Keynesian models suggest that bond prices should not necessarily rise due to such actions. The New Keynesian model incorporates future expectations, which may offset the immediate price increase of the bond for the investors. Investors may demand higher yields for holding bonds in the long term (\cite{JFJL19}). This discrepancy indicates that there is no consensus among different economic models regarding the effect of monetary policy on bond yields. By analyzing the impact of monetary policy surprises on bond yields within narrow windows around FOMC announcements, we can better understand the true impact of these surprises on bond prices. This paper focuses on how monetary policy influences treasury bond yields in the U.S., examining bond yields of various maturities through cross-sectional regression analysis. The research aims to provide an understanding of the different impacts during periods of Quantitative Easing and Quantitative Tightening, as well as during financial crises and non-crisis periods. 

This paper hypothesizes that unexpected monetary policy decisions announced during Federal Open Market Committee (FOMC) meetings significantly impact U.S. Treasury bond yields, with effects varying by context. During quantitative easing (QE) periods, short-term bond yields decrease due to increased demand from the Federal Reserve, while long-term bond yields remain unaffected. In financial crises, short-term bond yields significantly decrease due to heightened market volatility and risk aversion, while long-term bond yields stay stable. Combining QE and financial crises increases the sensitivity of 3-month and 10-year bond yields to economic and policy measures. Therefore, the study hypothesizes that financial crises significantly decrease short-term Treasury bond yields without significantly affecting long-term yields. QE alone does not significantly impact either short-term or long-term yields, but when combined with financial crises, it heightens the sensitivity of short-term yields to economic and policy changes.


\section{Literature Review}
\label{s:literature_review}

\subsection{Monetary Policy}

Monetary policy\footnote{Monetary policy is a set of actions available to a nation's central bank to achieve sustainable economic growth by adjusting the money supply.} influences bond yields significantly through interest rate adjustments and forward guidance, which is extensively documented in economic literature. Traditional monetary policy mechanisms, such as changes in the federal funds rate, influence short-term interest rates, which affect long-term bond yields, since higher short-term rates signal tighter monetary policy and potentially higher rates. Higher interest rates typically lead to higher bond yields, since investors would not prefer the existing bonds with lower fixed interest rates. Hence, investors demand higher yields to compensate for the expected increase in rates. 

Our study is guided by different literature, including Blinder’s (1990) research on ``The Federal Funds Rate and the Channels of Monetary Transmission" and D’Amico and King’s (2023) research on ``Past and Future Effects of the Recent Monetary Policy Tightening". In \cite{bernanke1990}, Blinder analyzed historical data on the federal funds rate and various macroeconomic indicators and found that the interest rate is a direct indicator of monetary policy actions, which records shocks to the supply of bank reserves. Recently, D’Amico and King (see \cite{ST23}) used the survey-augmented VAR\footnote{A statistical model that enhances a Vector Autoregression (VAR) -- which captures interdependencies among time series by using past values of all variables -- by integrating survey-based expectations.}  to study the effects of the current tightening cycle on output and inflation. They found that both expected and unexpected monetary policy changes impact economic conditions more quickly than traditional models predict, since the change of expectations leads to economic effects before implementing actual changes in the policy rate, suggesting a significant expectations channel. These findings underscore the importance of rate adjustment and market expectations in shaping bond yield movements, which guide us in analyzing the effect of monetary policy on bond yields.

\subsection{Quantitative Easing}

Quantitative easing (QE)\footnote{Quantitative easing (QE) is a form of monetary policy in which a central bank, like the United States Federal Reserve, purchases securities in the open market to reduce interest rates and increase the money supply.} has been pivotal in unconventional monetary policy, especially after the 2008 financial crisis.  Numerous studies have examined QE's impact on bond yields, with a consensus that QE lowers long-term yields by increasing the demand for long-term securities. Specifically, the Federal Reserve purchases long-term bonds and other financial assets, raising their prices due to heightened demand and lowering their yields. 

We reviewed some studies that confirm this statement, including Gagnon and others’ (2011) research on ``The Financial Market Effects of the Federal Reserve’s Large-Scale Asset Purchases" and Gagnon’s (2016) research on ``Quantitative Easing: An Underappreciated Success". Gagnon et al. (see \cite{JMJB11}) analyzed the situation when the Federal Reserve purchased substantial medium and long-maturity assets and explained how these purchases were implemented. They then presented evidence that these large-scale asset purchases led to meaningful and long-lasting reductions in longer-term interest rates on different securities by increasing the demand for securities. Additionally, Gagnon (\cite{gagnon2016}) reviewed studies of the estimated effects of QE bond purchases on 10-year yields using different methods, including time series and event studies. He then concluded that QE lowers bond yields significantly. These studies provide us with an overview of the impact of QE on T-bond yields.

There is an inverse relationship between bond yields and bond prices. Some studies suggest that QE also increases bond prices. After reviewing Todorov’s (\cite{todorov20}) research on ``Quantify the Quantitative Easing: Impact on Bonds and Corporate Debt Issuance", we found that the QE programme lowered bond yields and increased the prices and liquidity of corporate bonds eligible to be purchased substantially. This contradicts the New Keynesian model, which suggests that bond prices should not necessarily increase due to such actions. This contradiction gives us the incentive to study the influence of monetary policy on the government.

\subsection{The Inverse Relationship Between Bond Yield and Bond Price}

The inverse relationship between bond yields and prices is fundamental in finance and extensively documented in economic literature. When bond yields rise, bond prices fall, reflecting the lower present value of future cash flows. Conversely, when bond yields decrease, bond prices increase as the present value of future cash flows becomes more attractive. This relationship is crucial for understanding how interest rate changes impact bond investments.

We reviewed some literature, including Gagnon, J., Raskin, M., Remache, J., and Sack, B.'s research (see \cite{JMJB11}) on the financial market effects of the Federal Reserve’s large-scale asset purchases, and Krishnamurthy and Vissing-Jorgensen's work in 2011 (see \cite{AA11}). The effects of quantitative easing on interest rates: Channels and implications for policy. Brookings Papers on Economic Activity. Gagnon et al. (\cite{JMJB11}) explored this dynamic in the context of quantitative easing. They highlighted how large-scale asset purchases by central banks increase bond prices and decrease yields by boosting demand for these securities. Similarly, Krishnamurthy and Vissing-Jorgensen (\cite{AA11}) analyzed the effects of QE on interest rates. They found that increased demand for bonds due to QE leads to higher bond prices and lower yields, emphasizing the significant impact of central bank policies on bond markets.

\subsection{Financial Crises}

During financial crises, bond yields generally decrease as investors prefer safer assets. Central banks often respond with aggressive monetary policy actions, including large-scale asset purchases (QE), further depressing yields. Gilchrist, Lopez-Salido, and Zakrajsek (see \cite{SDE15}) show that during crises, the term premia on long-term bonds react strongly to policy shifts and indicate a significant market response to central bank interventions. Term premia refer to the extra yield investors seek for holding longer-term bonds. This yield compensates for the risks tied to long-term investments because markets are susceptible to changes in central bank policies, especially during crises. During such times, long-term real rates are more affected by unexpected monetary policy changes because of shifts in term premia. This observation challenges standard economic models. These models suggest that long-term real rates should be less affected by monetary policy after the period needed for nominal prices to adjust. Research by Hanson and Stein (\cite{SJ15}) highlights how long-term real rates are affected by monetary policy surprises, because they observed a stronger response during crisis periods due to changes in term premia. 

\subsection{Factors that Influence Bond Yields}

We reviewed different literatures, including \cite{SJ15} on Monetary policy and long-term real rates, \cite{DIT17} on risk aversion, sentiment and the cross-section of stock returns, \cite{MJ06} on investor sentiment and the cross-section of stock returns. We conclude the following three points:
\begin{enumerate}
\item[(1)] Monetary Policy: Central bank policy rates and expectations of future rates play a crucial role. For instance, changes in the federal funds rate and forward guidance from the Federal Reserve impact short-term and long-term yields differently. As seen in Hanson and Stein (\cite{SJ15}), monetary policy surprises can have intense and persistent effects on long-term real rates, primarily through changes in term premia. However, it may not be evident in the short term. 

\item[(2)] Inflation Expectations: Expected inflation is a key determinant of nominal bond yields. Higher expected inflation leads to higher nominal yields as investors demand compensation for losing purchasing power over time. This relationship is reflected in the break-even inflation rate, which is the difference between nominal and real yields.

\item[(3)] Market Sentiment and Risk Aversion: Investor behavior significantly impacts bond yields. During periods of high uncertainty or market stress, investors seek safe-haven assets like government bonds, which will increase demand and push yields down. Conversely, investors are more willing to take risks in stable periods, leading to higher bond yields as demand shifts towards riskier assets. Empirical studies support this relationship, showing that optimistic investor sentiment reduces risk aversion, boosting participation in riskier markets and elevating asset prices, including bond yields. \cite{DIT17}  highlights that investor sentiment strongly predicts bond risk premia, with a positive association between sentiment and market volatility. Baker and Wurgler (see \cite{MJ06})  also demonstrate that sentiment indicators predict bond risk premia effectively.
\end{enumerate}


\section{Regression and Data }

\subsection{Regression}

We consider the following regression model
\begin{equation*}
\begin{aligned}
\Delta Y_{i,t} = \beta_0 &+ \beta_1 \times \text{Crisis}_{t\_{crisis}} + \beta_2 \times \text{QE}_{t\_{QE}} + \beta_3 \times \text{Crisis}_{t\_{crisis}} \times \text{QE}_{t\_{QE}} \\
& + \beta_4 \times \text{FF}_t + \beta_5 \times \text{IR}_t + \epsilon.
\end{aligned}
\end{equation*}
Explanations for each term in the regression model are as follows:
\begin{itemize}
\item $\Delta Y_{i, t}$ represents the daily change in Treasury bond yields around the t, which is the FOMC announcement date. This could be calculated as $Y_{t+1} - Y_{t-1}$, where $Y_{t+1}$ is the treasury bond yield one day after the FOMC announcement date, and $Y_{t-1}$ is the treasury bond yield one day before the FOMC announcement date; $i$ represents different bond maturities, including $3$ months, $1$ year, $10$ years, and $30$ years.

\item $\text{Crisis}_{t\_{crisis}}$ is a  dummy variable indicating whether a financial crisis is present (1 if yes, 0 if no), and $t\_{crisis}$ represents the time period when there is financial crisis. 

\item $\text{QE}_{t\_{QE}}$ is a dummy variable indicating whether quantitative easing (QE) is being implemented (1 if yes, 0 if no), and $t\_{QE}$ represents the time period when there is QE.

\item $\text{Crisis}_{t\_{crisis}} \times \text{QE}_{t\_{QE}}$ is an interaction term between the crisis and QE dummy variables, capturing the combined impact on Treasury bond yields when both a financial crisis and QE are present.

\item $\text{FF}_t$ is a control variable that denotes for the federal fund rate at time $t$, the time of the FOMC announcement date.

\item $\text{IR}_t$ is a control variable that denotes for the inflation rate at time $t$, the time of the FOMC announcement date. 
\end{itemize}  

We want to analyze the difference in treasury bond yields in different conditions: 
\begin{enumerate}
\item In QE and financial crisis, 
\item Not in QE but in the financial crisis,
\item In QE but no financial crisis,
\item Not in QE and financial crisis.
\end{enumerate}
Thus, we have two dummy variables and one interaction term to differentiate the impacts. Besides, we will control for other macroeconomic variables that might influence bond yields, including inflation and federal funds rates. 

The inflation rate is calculated through the sticky price consumer price index (CPI), which is calculated from a subset of goods and services included in the CPI that change price relatively infrequently, so they are thought to incorporate expectations about future inflation to a greater degree than prices that change on a more frequent basis. One possible explanation for sticky prices could be the costs firms incur when changing prices. According to the regression above, the dependent variable represents the percentage change in the treasury bond yield. Thus, dealing with the inflation rate, we use the current month’s CPI level minus the last month’s CPI level and then divide by the current month's CPI level to get the inflation rate of the FOMC announcement dates. 
\begin{equation*}
\text{IR}_t = \frac{\text{CPI}_t - \text{CPI}_{t-1}}{\text{CPI}_{t}}
\end{equation*}

We consider inflation rates because of their direct impact on the real return on bonds. When inflation increases, the real yield decreases, prompting investors to demand higher nominal yields to maintain their returns in real terms and compensate for the loss of purchasing power (LiveWell, 2023)\footnote{Sneed, R. (2023, October 18). How does inflation affect bond yields? Livewell. https://livewell.com/finance/how-does-inflation-affect-bond-yields/.}.

Moreover, the federal funds rate level should be used as a control variable. According to the Federal Reserve, ``The federal funds rate is the interest rate at which depository institutions trade federal funds with each other overnight". As a primary monetary policy tool, changes in the federal funds rate influence short-term interest rates and, consequently, longer-term treasury yields. It directly affects the cost of borrowing and the overall economic activity. Specifically, an increase in the federal funds rate makes borrowing more expensive, which helps to cool down an overheating economy, leading to higher yields on treasury bonds as demand for bonds decreases. Conversely, decreasing the rate aims to stimulate economic activity by lowering borrowing costs, resulting in lower bond yields, because demand for bonds increases (Federal Reserve Board, n.d.)\footnote{Principles for the Conduct of Monetary Policy. (n.d.). https://www.federalreserve.gov/monetarypolicy/principles-for-the-conduct-of-monetary-policy.htm}. According to the regression model, we use the federal funds rate level as the control variable because there is no evident difference in the FOMC announcement dates, the day before and after those dates. 

\begin{table}[!h]
\centering
\caption{Statistics Summary}
\label{tab:table1}
\begin{tabular}{@{}ccccc@{}}
\toprule
 & \textbf{3-month} & \textbf{1-year}  & \textbf{10-year} & \textbf{30-year}  \\ \midrule
\vspace{8pt}
\textbf{Size} & 6135  &  6135 & 6135 & 6135 \\
\vspace{8pt} \textbf{Mean} & 1.79  & 1.98  & 3.27 & 3.87 \\
\vspace{8pt} \textbf{St. Deviation} & 1.96 & 1.89 & 1.31 & 1.19 \\
\vspace{8pt} \textbf{Minimum} & 0.00  & 0.04 & 0.52 & 0.99 \\
\vspace{8pt} \bf{25\%} & 0.10  & 0.26 & 2.19 & 2.95 \\
\vspace{8pt} \bf{50\%} & 1.07  & 1.36 & 3.21 & 3.90 \\
\vspace{8pt} \bf{75\%} & 3.00  & 3.34 & 4.28 & 4.82 \\
\vspace{2pt} \textbf{Maximum} & 6.42  & 6.44 & 6.79 & 6.75 \\
\bottomrule
\end{tabular}
\end{table}

\subsection{Data}

All data were sourced from the Federal Reserve Economic Data (FRED) database. The dataset spans from January 2000 to July 2024, encompassing all Federal Open Market Committee (FOMC) announcement dates, totaling 196 announcements. In our data, with information limitations, 191 out of 196 FOMC announcement dates are included. The regression model includes variables such as the intercept (constant), indicators for financial crises (Crisis), quantitative easing (QE), the interaction term between Crisis and QE, the federal funds rate, and the inflation rate. We conducted four separate regressions, each corresponding to different maturities of Treasury bonds: 3-month, 1-year, 10-year, and 30-year Treasury bonds.

\begin{table}[!h]
\centering
\caption{Basic regression information}
\label{tab:table2}
\begin{tabular}{@{}cc@{}}
\toprule
\vspace{8pt}
\textbf{Dep. Variable} & \text{Yield\_Change} \\
\vspace{8pt} \textbf{Model} & \text{OLS} \\
\vspace{8pt} \textbf{Mothod} & \text{Least Square} \\
\vspace{8pt} \textbf{Date} & \text{Tue, 23 Jul 2024} \\
\vspace{8pt} \textbf{Time} & 18:52:27 \\
\vspace{2pt} \textbf{No. Observations} & 191 \\
\bottomrule
\end{tabular}
\end{table}


\section{Result}

\subsection{Regression Results for 3-Month Treasury Bond}

The intercept represents the average change in 3-month Treasury bond yields around FOMC dates when there is no financial crisis, no QE, and control variables at their mean values. The coefficient is statistically significant ($p$-value $< 0.1$), indicating a significant average decrease in 3-month Treasury bond yields without the main effects and controls. The crisis variable is highly significant ($p$-value $< 0.1$), indicating that financial crises significantly decrease 3-month Treasury bond yields by approximately $0.2102$ percentage points, holding other factors constant. The QE variable is not statistically significant ($p$-value $> 0.1$), suggesting that QE alone does not have a significant direct impact on $3$-month Treasury bond yields around FOMC dates in this model. The interaction term between Crisis and QE is highly significant ($p$-value $< 0.1$), indicating that the combined effect of a financial crisis and QE significantly increases $3$-month Treasury bond yields by approximately $0.2388$ percentage points, holding other factors constant. For both control variables, the federal funds rate and the inflation rate are not statistically significant. This result indicates that the $10$-year treasury bond yields around the FOMC dates and the monetary policy surprises are unaffected by them.

\begin{table}[h!]
\centering
\caption{Regression results for 3-month treasury bond}
\label{tab:table3}
\begin{tabular}{@{}ccccccc@{}}
\toprule
 & \textbf{Coef} & \textbf{Std err}  & \textbf{t} & $\bf{P > |t|}$ & \bf{[0.025}  & \bf{0.975]} \\ \midrule
\vspace{8pt}
\text{Const} & $- 0.0281$  & 0.011 & $-2.443$ & 0.015 & $- 0.051$ & $-0.005$ \\
\vspace{8pt} \text{Crisis} & $-0.2102$  & 0.026  & $-8.096$ & 0.000 & $- 0.261$ & $-0.159$ \\
\vspace{8pt} \text{QE} & 0.0226 & 0.017 & 1.350 & 0.179 & $-0.010$ & 0.056 \\
\vspace{8pt} \text{Crisis\_QE} & 0.2388  & 0.052 & 4.598 & 0.000 & 0.136 & 0.341 \\
\vspace{8pt} \text{FF} & 0.0019  & 0.004 & 0.504 & 0.615 & $- 0.005$ & 0.009 \\
\vspace{2pt} \text{IR} & $-0.0393$  & 0.105 & $-0.373$ & 0.709 & $- 0.247$ & 0.169 \\
\bottomrule
\end{tabular}
\end{table}

\subsection{Regression Results for 1-Year Treasury Bond}

The intercept represents the average change in 1-year Treasury bond yields around FOMC dates when there is no financial crisis, not in QE periods, and control variables at their mean values. The crisis variable is statistically significant ($p$-value $< 0.1$), indicating that financial crises significantly decrease 1-year Treasury bond yields by approximately $0.0469$ percentage points, holding other factors constant. The QE variable is not statistically significant ($p$-value $> 0.1$), suggesting that QE alone does not have a significant direct impact on 1-year Treasury bond yields around FOMC dates in this model. The interaction term between Crisis and QE is not statistically significant ($p$-value $> 0.1$), suggesting that the combined effect of a financial crisis and QE on 1-year Treasury bond yields is not statistically significant in this model. The federal funds rate is statistically significant ($p$-value $< 0.1$), indicating that an increase in the federal funds rate is associated with a decrease in 1-year Treasury bond yields by approximately $0.0083$ percentage points, holding other factors constant. The inflation rate variable is not statistically significant ($p$-value $> 0.1$), indicating that changes in the inflation rate do not have a significant direct impact on 1-year Treasury bond yields around FOMC dates in this model.

\begin{table}[h!]
\centering
\caption{Regression results for 1-year treasury bond}
\label{tab:table4}
\begin{tabular}{@{}ccccccc@{}}
\toprule
 & \textbf{Coef} & \textbf{Std err}  & \textbf{t} & $\bf{P > |t|}$ & \bf{[0.025}  & \bf{0.975]} \\ \midrule
\vspace{8pt}
\text{Const} & $- 0.0024$  & 0.008 & $-0.293$ & 0.770 & $- 0.018$ & 0.014 \\
\vspace{8pt} \text{Crisis} & $-0.0469$  & 0.018  & $-2.585$ & 0.011 & $- 0.083$ & $-0.011$ \\
\vspace{8pt} \text{QE} & 0.0012 & 0.012 & 0.100 & 0.921 & $-0.022$ & 0.024 \\
\vspace{8pt} \text{Crisis\_QE} & 0.0158  & 0.036 & 0.434 & 0.665 & $-0.056$ & 0.087 \\
\vspace{8pt} \text{FF} & $-0.0083$  & 0.003 & $-3.223$ & 0.002 & $- 0.013$ & $-0.003$ \\
\vspace{2pt} \text{IR} & $-0.0345$  & 0.074 & $-0.468$ & 0.640 & $- 0.180$ & 0.111 \\
\bottomrule
\end{tabular}
\end{table}

\subsection{Regression Results for 10-Year Treasury Bond}

The intercept represents the average change in 10-year Treasury bond yields around FOMC dates when there is no financial crisis, not in QE periods, and control variables at their mean values. The crisis variable is not statistically significant ($p$-value $> 0.1$), suggesting that financial crises alone do not have a significant impact on $10$-year Treasury bond yields around FOMC dates in this model. The QE variable is not statistically significant ($p$-value $> 0.05$), indicating that QE alone does not have a significant direct impact on 10-year Treasury bond yields around FOMC dates in this model. The interaction term between Crisis and QE is statistically significant ($p$-value $< 0.1$), indicating that the combined effect of a financial crisis and QE significantly decreases 10-year Treasury bond yields by approximately $0.1457$ percentage points, holding other factors constant. For both control variables, the federal funds rate and the inflation rate are not statistically significant. This result indicates that the 10-year treasury bond yields around the FOMC dates, the monetary policy surprises, are not affected.

\begin{table}[h!]
\centering
\caption{Regression results for 10-year treasury bond}
\label{tab:table5}
\begin{tabular}{@{}ccccccc@{}}
\toprule
 & \textbf{Coef} & \textbf{Std err}  & \textbf{t} & $\bf{P > |t|}$ & \bf{[0.025}  & \bf{0.975]} \\ \midrule
\vspace{8pt}
\text{Const} & $- 0.0112$  & 0.014 & $-0.802$ & 0.423 & $- 0.039$ & 0.016 \\
\vspace{8pt} \text{Crisis} & 0.0252 & 0.031 & 0.801 & 0.424 & $- 0.037$ & 0.087 \\
\vspace{8pt} \text{QE} & 0.0288 & 0.020 & 1.421 & 0.157 & $-0.011$ & 0.069 \\
\vspace{8pt} \text{Crisis\_QE} & $-0.1457$  & 0.063 & $-2.318$ & 0.022 & $-0.270$ & $-0.022$ \\
\vspace{8pt} \text{FF} & $-0.0072$  & 0.004 & $-1.605$ & 0.110 & $- 0.016$ & 0.002 \\
\vspace{2pt} \text{IR} & $-0.1036$  & 0.128 & $-0.812$ & 0.418 & $- 0.355$ & 0.148 \\
\bottomrule
\end{tabular}
\end{table}

\subsection{Regression Results for 30-Year Treasury Bond}

The intercept represents the average change in $30$-year Treasury bond yields around FOMC dates when there is no financial crisis, not in quantitative easing periods, and control variables at their mean values. The intercept’s coefficient is not statistically significant, indicating no significant average change in $30$-year Treasury bond yields without the main effects and controls. The crisis variable is not statistically significant ($p$-value $> 0.1$), suggesting that financial crises alone do not have a significant impact on $30$-year Treasury bond yields around FOMC dates in this model. The QE variable is statistically significant ($p$-value $< 0.1$), indicating that QE significantly positively impacts $30$-year Treasury bond yields around FOMC dates. QE periods are associated with an increase in $30$-year Treasury bond yields by approximately $0.0362$ percentage points, holding other factors constant. The interaction term between Crisis and QE is not statistically significant ($p$-value $> 0.1$), suggesting that the combined effect of a financial crisis and QE on $30$-year Treasury bond yields is not statistically significant in this model.  For both control variables, the federal funds rate and the inflation rate are not statistically significant. This result indicates that the $30$-year treasury bond yields around the FOMC dates and the monetary policy surprises are unaffected by them. 

\begin{table}[h!]
\centering
\caption{Regression results for 30-year treasury bond}
\label{tab:table6}
\begin{tabular}{@{}ccccccc@{}}
\toprule
 & \textbf{Coef} & \textbf{Std err}  & \textbf{t} & $\bf{P > |t|}$ & \bf{[0.025}  & \bf{0.975]} \\ \midrule
\vspace{8pt}
\text{Const} & $- 0.0146$  & 0.012 & $-1.170$ & 0.244 & $- 0.039$ & 0.010 \\
\vspace{8pt} \text{Crisis} & 0.0069 & 0.028 & 0.246 & 0.806 & $- 0.049$ & 0.062 \\
\vspace{8pt} \text{QE} & 0.0362 & 0.018 & 1.994 & 0.048 & 0.000 & 0.072 \\
\vspace{8pt} \text{Crisis\_QE} & $-0.0599$ & 0.056 & $-1.067$ & 0.287 & $-0.171$ & 0.051 \\
\vspace{8pt} \text{FF} & $-0.0037$  & 0.004 & $-0.929$ & 0.354 & $- 0.012$ & 0.004 \\
\vspace{2pt} \text{IR} & $-0.0518$  & 0.114 & $-0.454$ & 0.650 & $- 0.277$ & 0.173 \\
\bottomrule
\end{tabular}
\end{table}

\subsection{Short Term Analysis}

According to the regression results above for 3 months and 1 year treasury, we can see that financial crises have a significant negative impact on short-term yields. During financial crises, the 3-month Treasury bond yield decreases by $0.2102$ percentage points, and the 1-year Treasury bond yield decreases by $0.0469$ percentage points. The Federal Reserve often responds to financial crises by lowering short-term interest rates to stimulate the economy. During crises, investors expect prolonged economic difficulties and prefer the liquidity provided by short-term securities. This behavior leads to a decrease in short-term yields as investors move away from riskier assets. In uncertain times, the demand for liquid and safe assets increases, further decreasing short-term yields. The reason why QE is not significant may be that QE mainly targets long-term interest rates more directly. Short-term yields are more influenced by short-term interest rate policies and immediate economic conditions. The combined effect of crises and QE significantly increases the 3-month yield, suggesting heightened sensitivity to immediate economic and policy measures. 

\subsection{Long Term Analysis}

According to the regression results above for 10-year and 30-year treasuries, the impact of financial crises and QE on long-term yields has a tiny difference. The data indicate that 10-year Treasury bond yields do not exhibit significant changes during financial crises, showing a neutral impact. The QE periods also do not significantly affect 10-year Treasury bond yields on their own. However, the combined effect of a financial crisis and QE leads to a significant decrease in 10-year Treasury bond yields by approximately $0.1457$ percentage points. This decrease reflects heightened market sensitivity to economic distress and policy intervention during crises.

For 30-year Treasury bonds, financial crises alone do not significantly impact yields. Also, QE doesn't have an obvious effect, as it only increases Treasury bond yields by $0.0362$ percentage points. This suggests that long-term bonds, such as economic conditions and inflation, are not very sensitive to QE. The interaction of financial crises and QE does not show a significant combined effect on 30-year yields, indicating that each factor independently influences long-term yields. In times of economic uncertainty, investors demand higher risk premiums to hold long-term bonds, offsetting the rate-lowering effect of QE. Secondly, during financial crises, the increased demand for safe-haven assets drives up long-term bond prices and lowers yields; however, this increased demand might have already been anticipated by the market, thus limiting the impact. Lastly, when the market expects an economic downturn, investors might anticipate the central bank lowering short-term rates, leading them to flock to long-term bonds, which reduce long-term yields and create an inverted yield curve.

The preferences of pension funds and larger institutions for long-term treasuries also help explain the regression results. These institutions prefer long-term treasuries for liability matching, regulatory compliance, and stable income generation. Their consistent demand for long-term bonds stabilizes yields, even during economic uncertainty and aggressive monetary policy periods. This stability is crucial for these institutions. It helps them preserve capital and ensure predictable returns over extended periods.

We also pay attention to the impact of market segmentation on bond yields. The persistent demand from institutional investors for long-term treasuries reduces the immediate effects of QE and financial crises on these bonds. These investors are less sensitive to short-term economic policies. They are more focused on long-term economic outlooks. Their behavior helps explain why 30-year yields remain independent from short-term market shocks.

\subsection{Suggestions for Portfolio Managers}

According to the result of the regression model, monetary policy influences short-term and long-term T-bond yields differently. Financial crises decrease short-term T-bond yields significantly, but the impact of QE is not obvious. Due to this situation, portfolio managers might increase the allocation to T-bonds during financial crises, since lower yields during that period indicate higher bond prices, which is beneficial in a declining yield environment. Managers can also stay vigilant about economic indicators and signals that could predict financial crises. Although QE does not significantly impact short-term yields based on the model, it is still advantageous for managers to diversify their portfolios to mitigate risks associated with other factors that might influence yields, including GDP growth and government fiscal policy. 

On the contrary, neither financial crises nor QE independently influence long-term T-bond yield, while there is a negative combination effect of QE and financial crisis for 10-year treasury bonds. Therefore, managers can consider diversification, such as including bonds with a mix of durations in their portfolio, which helps to reduce the risk given the return. Additionally, although the model suggests a limited direct impact from QE and financial crises on long-term yields, other factors may still play a role in shaping bond yields, so it is beneficial for managers to pay attention to other economic indicators to make informed decisions. The negative combination effect of QE and financial crises on 10-year treasury bond yields suggests the advantage of hedging strategies to protect against potential declines in long-term bond prices, including using options or futures to hedge their positions. 

When the yield curve is inverted, portfolio managers probably invest more in short-term bonds; when a normal yield curve is presented, managers might invest more in long-term bonds, given lower reinvestment risk (Howard, 2023)\footnote{Why Go Long When Short-Term Bonds Yield More? (2023). Schwab Brokerage. https://www.schwab.com/learn/story/why-go-long-when-short-term-bonds-yield-more}.

\section{Conclusion}

This study examines the effects of monetary policy and financial crises on the U.S. Treasury bond yields, focusing on the impacts of QE and the financial crisis.  By analyzing the effects of monetary policy surprises on bond yields within narrow windows around FOMC announcements, this study isolates the impact of monetary surprises, offering a picture of how unexpected monetary policy decisions influence the government bond market. The findings resonate with our hypotheses: financial crises significantly decrease short-term Treasury yields as investors seek safe-haven assets, aligning with existing literature on market reactions during economic turmoil. Conversely, QE does not affect bond yield significantly. For long-term Treasury bonds, financial crises show no significant impact, likely due to stable institutional demand, while QE significantly influences long-term yields, particularly during crises.

These insights are crucial for portfolio managers to manage interest rate risks and make informed investment decisions. Our research contributes to the literature by bridging theoretical predictions with empirical evidence, highlighting the complex interactions between economic factors and bond markets.

\bibliographystyle{plain}
\bibliography{reference}

\appendix
\section{Appendix}
\subsection{Financial Crisis}

\begin{figure}[h!]
\centering
\includegraphics[width=.48\textwidth]{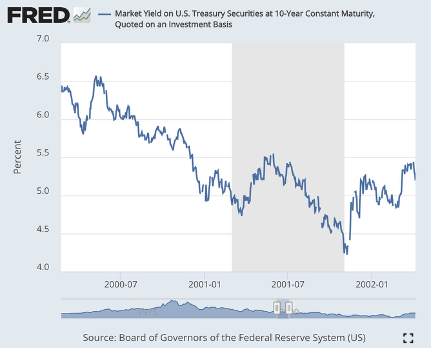}
\caption{Financial Crisis from Feb 28, 2001 to Nov 1, 2001.}
\label{fig:figure1}
\end{figure}

\begin{figure}[h!]
\centering
\includegraphics[width=.48\textwidth]{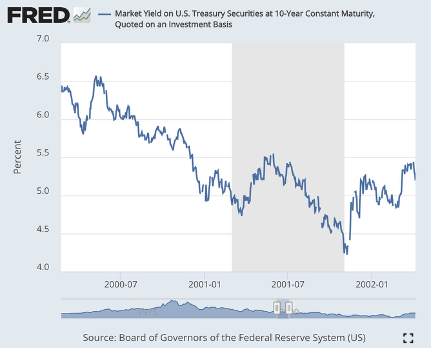}
\caption{Financial Crisis from Dec 3, 2007 to June 1, 2009.}
\label{fig:figure2}
\end{figure}

\begin{figure}[h!]
\centering
\includegraphics[width=.48\textwidth]{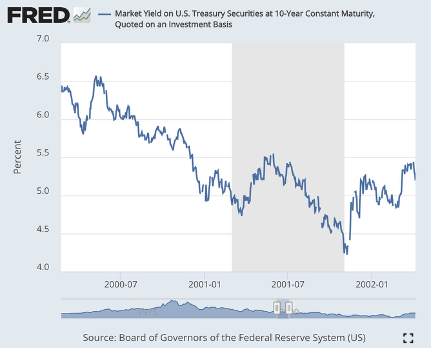}
\caption{Financial Crisis from Jan 30, 2020 to Apr 1, 2020.}
\label{fig:figure3}
\end{figure}

\subsection{Quantitative Easing Period}

\begin{table}[!h]
\centering
\caption{Quantitative Easing Period from 2000 to 2024}
\label{tab:table7}
\begin{tabular}{@{}cc@{}}
\toprule
 & \textbf{QE Period (2000-2024)} \\ \midrule
\vspace{8pt} \text{QE1} & \text{November 2008 - March 2010} \\
\vspace{8pt} \text{QE2} & \text{November 2010 - June 2011} \\
\vspace{8pt} \text{QE3} & \text{September 2012 - October 2014} \\
\vspace{1pt} \text{QE4} & \text{March 2020 - March 2022} \\
\bottomrule
\end{tabular}
\end{table}

\end{document}